# SCAview: Interactive cohort visualization and exploration for spinocerebellar ataxias using synthetic cohort data

Philipp Wegner[1,2,*], Sebastian Schaaf[1], Mischa Uebachs[3], Marcus Grobe-Einsler[1,4], Sumit Madan[2,6], Thomas Klockgether[1], Juliane Fluck[5,6], Jennifer Faber[1,4]

(1) German Center for Neurodegenerative Diseases (DZNE), Bonn, Germany, https://ror.org/043j0f473 (2) Department of Bioinformatics, Fraunhofer Institute for Algorithms and Scientific Computing (SCAI), Sankt Augustin 53754, Germany, https://ror.org/00trw9c49 (3) DRK Kamillus Klinik Asbach Neurology, 53567 Asbach, Germany, https://ror.org/009gj6k64

(4) Department of Neurology, University Hospital Bonn, Bonn, Germany

(5) https://orcid.org/0000-0003-1379-7023, ZB Med, Köln 50931, Germany, https://ror.org/0259fwx54

(6) University of Bonn, Bonn 53115, Germany https://ror.org/041nas322

*To whom correspondence should be addressed.

## Abstract

**Motivation:** Visualization of data is a crucial step to understanding and deriving hypotheses from clinical data. However, for clinicians, visualization often comes with great effort due to the lack of technical knowledge about data handling and visualization. The application offers an easy-to-use solution with an intuitive design that enables various kinds of plotting functions. The aim was to provide an intuitive solution with a low entrance barrier for clinical users. Little to no onboarding is required before creating plots, while the complexity of questions can grow up to specific corner cases. To allow for an easy start and testing with SCAview, we incorporated a synthetic cohort dataset based on real data of rare neurological movement disorders: the most common autosomal-dominantly inherited spinocerebellar ataxias (SCAs) type 1, 2, 3, and 6 (SCA1, 2, 3 and 6).

**Methods:** We created a Django-based backend application that serves the data to a React-based frontend that uses Plotly for plotting. A synthetic cohort was created to deploy a version of SCAview without violating any data protection guidelines. Here, we added normal distributed noise to the data and therefore prevent re-identification while keeping distributions and general correlations.

**Results:** This work presents SCAview, an user-friendly, interactive web-based service that enables data visualization in a clickable interface allowing intuitive graphical handling that aims to enable data visualization in a clickable interface. The service is deployed and can be tested with a synthetic cohort created based on a large, longitudinal dataset from observational studies in the most common SCAs.

**Availability:** http://idsn.dzne.de - In order to get a user and password please contact scaview@dzne.de

**Code:** https://github.com/DZNE-ataxia-research/SCAview

**Contact:** scaview@dzne.de

**Supplementary information:** Supplementary data is appended.

## Introduction

Visualizing data can significantly contribute to understanding and active interpretation of the data at hand. However, the preparation of data visualization is often tedious and demanding, as it involves data handling and pre-processing, for which certain technical skills are required. Moreover, to finally plot the data, one needs the ability to work with the appropriate libraries such as Pyplot (https://matplotlib.org/stable/tutorials/introductory/pyplot.html) or Seaborn (https://seaborn.pydata.org/). Pre-built software exists, such as the i2b2 analysis tool [1]. i2b2 is a powerful solution that allows filtering for patient sets and plotting various clinical data. However, extensive technical knowledge is already required for installation and usage, which is a significant barrier for non-technical researchers to use such tools for initial data analysis. A visualization tool should meet the everyday needs of clinicians, in particular a quick and easy visualization of data. In addition to ease of use, the availability of datasets is often limited. Especially for rare diseases, individual centers usually have only limited data available, and access to multicentric data is rather limited due to particular data protection regulations. As a use case and for the implementation, cohort data of ataxia studies and longitudinal data from European and US natural history studies of spinocerebellar ataxia type 1, 2, 3, and 6, were used. This outstanding large dataset found the basis for a synthetic cohort, that is integrated into the public SCAview web service allowing a broad community of researchers and clinicians to browse a reasonable number of rare disease cases on their own. SCAview was developed in close cooperation with clinicians (non-specialists as well as specialists for ataxias) and their feedback was continuously included in the tool development. In summary, SCAview's goal is to provide a lightweight open-source visualization tool that can be deployed behind the firewall of protected areas[1] of a hospital or research institute with a special focus

---

[1] Such as virtual machines that align with the clinic's data protection guidelines

on ease of use and the opportunity to browse a large synthetic dataset of rare neurological diseases.

## Methodology and Implementation

The SCAview visualization tool is based on a data and a visualization service. The data service acts as a backend for processing and providing the data, while the visualization service functions as a frontend interacting with the user. The backend is developed as a Django (https://www.djangoproject.com/) application. The frontend uses React (https://reactjs.org/) and the React version of Plotly (https://plotly.com/javascript/react/) to create interactive plots. The data itself as well as the user sessions are persisted in a RedisDB (https://redis.io/), which allows instant data access. The tool stores every plot and its settings in a user session, allowing users to seamlessly continue working with their plots even after closing and re-opening the browser. Adopting Plotly, the application implements scatter plots, histograms, bar plots as well as timeline plots. The whole stack of components used in SCAview has little performance requirements as it was tested on standard personal laptops and ran smoothly throughout various demonstrations.

The visualizer was intentionally created to visualize and analyze an ataxia dataset that was assembled from various different sources comprising SCAregistry [2], EUROSCA [3], CRC-SCA [4], and RISCA [5]. This dataset is harmonized with respect to the individual data variables and contains 115 variables that were measured over up to 9 visits in a total of 1,417 patients. Here, we define a visit as a patient coming to the hospital/medical institution/research center in the scope of the study. A visit includes standardized clinical assessments of commonly used scales to assess the severity of the core symptom ataxia, neurological symptoms other than ataxia as well as information on the disease stage. Characterizing data such as demographic and genetic information are assessed at baseline. Multiple visits at the hospital are merged into one if they occur in a time span of less than 28 days.

The viewer is accessible via a public URL restricted by a password. In order to address privacy concerns as well as protect patients' personal information, a synthetic cohort was derived from the original cohort that preserves global linear correlations, but changes critical data points sufficiently to mask the original values and thereby reduce the risk of re-identification to a negligible minimum. The synthetic cohort was generated patient-wise, where a synthetic value for a variable $X$ (one measurement for a particular patient at one visit) of the original cohort is derived via the model $f(X) = X + \epsilon$, where $\epsilon \sim N(0, v)$, $v = Var(X) * b$, $b \in (0, 1]$, where $Var(X)$ is the empirical variance of all data points available for the variable $X$. Further, note that the term $b$ is a scalar necessary to regularize the variances of synthetic variables (**Supplementary Text 1**).

Vividly speaking, this method adds normal distributed "noise" to the original data. This method is applied throughout the entire dataset for each patient and all critical variables[2] corresponding to them. For categorical variables, such as test scores, the method is applied equally followed by a mapping step that maps the output $f(X)$ to the respective category. For instance, if some test score has discrete values in $\{1, 2, 3, 4, 5\}$ and $X = 5$, $f(X) = 4.31$, then $map(f(X)) = 4$. The described method ensures preserving the mean of each variable since $E[f(X)] = E[X + \epsilon] = E[X] + E[\epsilon] = E[X] + 0$. Moreover, the linear correlation of two variables $X, Y$ measured by the Pearson correlation coefficient $\rho_{X,Y} = \frac{cov[X,Y]}{\sigma_X \sigma_Y}$ is preserved by the above model up to a scaling factor depending on $b$. It holds that $\rho_{X,Y} = (1 + b)\rho_{\tilde{X},\tilde{Y}}$, where $X, Y$ the original variables of the dataset and $\tilde{X}, \tilde{Y}$ the corresponding synthetic variables (**Supplementary Text 2**). That relationship between the correlation coefficients reveals that in order to preserve linear correlations as good as possible, $b$ should be chosen as small as possible. However, smaller values for $b$ add less noise to the original dataset which might alter the variables not enough. Hence, the exact value for $b$ needs to be selected based on the extent to which the data needs to be noised. To the best of our knowledge, the applied method is considered to alter the data enough such that no re-identification of patients is possible with reasonable efforts.

## Results

The frontend of SCAview allows the exploration of clinical data in an interactive viewer that enables data visualization and exploration of a synthetic cohort derived from a large dataset of standardized clinical assessments in rare neurological diseases. The backend handles all of the data logic, which comprises sub-filtering of data, adjusting the data to different plot types, storing and querying sub-groups as well as session management. With that backend functionality implemented the frontend is able provide the following features.

*General filtering of items.* The user can either select the data to be filtered patients-wise or visits-wise. The patient-wise viewing includes all patient visits where at least during one visit the value suits the selected boundaries ("any"), whereas the visits-wise viewing selects only the exact visit that matches the current filter.

*Plot types.* The tool allows plotting four different types of plots namely histogram, scatter plot, bar plot, and timeline plot. Certain plot types are limited to a particular type of data e.g. timeline plots can only display data with a time dimension such as age (related to the variables *date of birth and visit date)* or the *reported disease duration.* All plots created by the user are displayed centered in a grid-like arrangement, selectable by clicking. On the right-hand side of the application, the plot type, data variables to be plotted, and further options, such as for linear regression fits, are located.

*Subgroup definition.* Subgroups can be defined graphically or via a filter panel. On the left-hand side, the user can set filters, which allows them to manually edit stratification parameters to define subgroups. Furthermore, subgroups can be defined graphically, by drawing a rectangle on a selected area in a plot that outlines the desired values. The latter, in particular, enables setting filters intuitively based on data distributions. Several filter settings can be combined and in addition, any filter or filter set can be saved as a tagged subgroup. Thus, further analyses of subgroups of interest can be performed.

*Easy reset.* Furthermore, buttons located at the top allow resetting of the current session, which deletes all plots, filters, and named subgroups and clears the whole view (**Figure 1**).

Finally, the software is currently deployed with a synthetic cohort in order to protect data privacy while still allowing the exploration of realistic data. This cohort is equal in the number of patients to the original spinocerebellar ataxia (SCA) dataset, hence it includes 1,417 unique patients with 255,027 data points, ready to be explored by ataxia researchers and clinicians. General distributions and correlations were visually inspected by ataxia experts and compared to the corresponding plots in the real dataset.

---

[2] Those variables that would allow re-identification of a patient

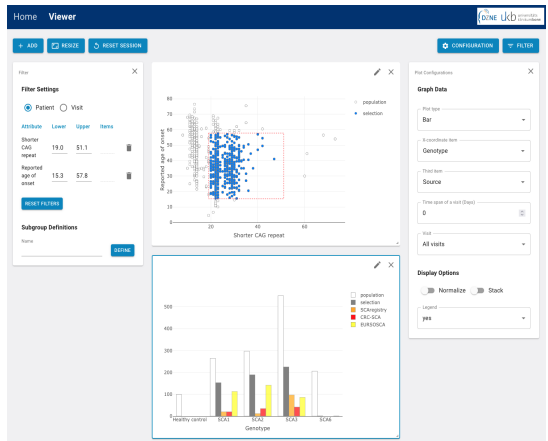

**Figure 1**: SCAview interface with two plots in parallel view

## Discussion and Future Work

SCAview is a browser-based visualization tool, that provides an easy-to-use interface that can, in a few steps, create plots and combinations of plots with great expressive power, enabling a convenient way of highly interactive data exploration. Various plot types and intuitive creation of subgroups allow visualization of particular interest allowing one to gain a deeper knowledge and generate hypotheses. The low entrance barrier with almost no technical skills necessary for the implementation and usage are a big advantage of SCAview, in contrast to tools with complex interfaces that usually have a steeper learning curve such as i2b2 [1]. The creation of a synthetic cohort based on real data of more than 1,400 patients suffering from rare neurological movement disorders, namely spinocerebellar ataxia type 1, 2, 3, and 6, was necessary due to data protection. However, the chosen approach preserves means and relations, thus allowing to study almost realistic dataset. SCAview with integrated analysis of the original as well as the synthetic cohort has been successfully introduced as a resource in the "Ataxia Global Initiative" *Uebachs et al.* [6], a worldwide platform for clinical research. The acceptance in the ataxia community will show, whether SCAview is a useful tool for clinical research. We chose a rare neurological disease as a use case since access and availability to data in rare diseases is in particular challenging. However, the tool functions implemented in SCAview are able to handle data from various sources and can easily be modified for general-purpose visualizations.

Finally, SCAview nevertheless has some limitations. For example, it allows currently only a linear regression of data. We plan to include further functionalities in a next version, e.g. an attribute workbench that would enable users to craft custom variables or to add other methods, e.g. data interpolation, in a user-friendly way. Moreover, the next steps in increasing SCAviews capabilities would include the improvement of already existing functionality like adding more options for data fitting. Finally, the quality of the synthetic cohort could be improved by applying more sophisticated methods to create synthetic patient data such as GANs (Generative Adversarial Networks) or VAEs (Variational Autoencoders).

## Funding

This project has received funding from the German Federal Ministry of Education and Research (BMBF) as part of the program "i:DSem—Integrative Data Semantics in the Systems Medicine", project number 031L0029 [A-C].

## Author contributions

Jennifer Faber, Mischa Uebachs, Juliane Fluck, Sumit Madan, Marcus Grobe-Einsler, and Thomas Klockgether continuously tested the tool and provided valuable feedback, and contributed to the conceptualization of the project. Mischa Uebachs, Sebastian Schaaf, and Philipp Wegner contributed to developing the underlying software of SCAview.

# Supplementary File: SCAview: Interactive cohort visualization and exploration for spinocerebellar ataxias using synthetic cohort data

## Supplementary Text 1 - Deriving a suitable b value for the noise-adding model:

We used the following model to generate a synthetic cohort based on real patient data.

$$f(X) = X + \epsilon, \text{ where } \epsilon \text{ is } N(0, Var(X) * b) \qquad (1)$$

Where $X$ is the original value for the variable to be noised, $\epsilon$ a normal distributed variable drawn from $N(0, Var(X) * b)$, $Var(X)$ the variance of $X$ and $b$ a regularizing scalar.

In order to generate a synthetic cohort that mimics the original data as well as possible, one has to adjust the b-value for the above-mentioned model (1).

Adjusting the b-value has a direct effect on the distribution of the synthetic data, and consequently on the similarity to the data distribution in the original data set. Note that the $b$ value is determined per patient and per variable to meet certain criterias described below. Supplementary Figure 1 shows kernel density estimated plots for an exemplarily variable (SARASUM) and different values of b (including the iteratively determined one implemented in the current version) as well as the original data set. Note that for each variable X, the corresponding b value is determined in an iterative process that initially samples new variables with b=1 and validates the new synthetic variable f(X) whether the synthetic value is plausible with respect to certain goodness criteria (e.g. max. 2.5% deviation from original value for variable SARASUM) or not. If those criteria, such as plausible value ranges, are not met, the b value is scaled by some step size (0.9 for float and 0.85 for integer in the current version), and a new synthetic variable is sampled. This is repeated until the goodness criteria are met.

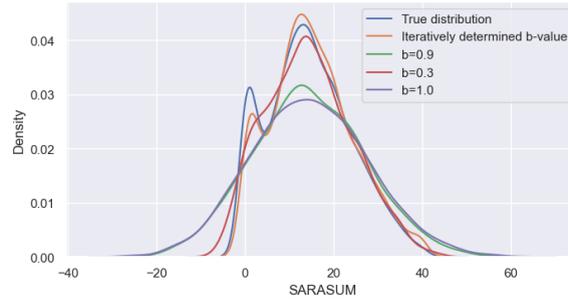

**Supplementary Figure 1:** Synthetic data for different b values

## Supplementary Text 2:

The following depicts how the applied method preserves linear correlation measured by the Pearson correlation coefficient.

Let $\tilde{X} = f(X)$ a synthetic value for a real value $X$ (analog for $Y$), $\rho_{X,Y}$ the Pearson correlation coeff. for $X$ and $Y$ and $\epsilon$ and $b$ as in Supplementary Text 1.

$$Var[\tilde{X}] = Var[X] + Var[\epsilon] = Var[X] + Var[X] * b$$
$$\Rightarrow \sigma_{\tilde{X}} = \sqrt{(1+b)Var[X]}$$

Putting that into the definition of the Pearson correlation coefficient allows the following deduction.

$$\rho_{\tilde{X},\tilde{Y}} = \frac{Cov[\tilde{X},\tilde{Y}]}{\sigma_{\tilde{X}}\sigma_{\tilde{Y}}} = \frac{Cov[\tilde{X},\tilde{Y}]}{\sqrt{(1+b)}\sigma_X \sqrt{(1+b)}\sigma_Y} = \frac{1}{(1+b)} \frac{Cov[\tilde{X},\tilde{Y}]}{\sigma_X \sigma_Y}$$

$$Cov[\tilde{X},\tilde{Y}] = Cov[X + \epsilon_X, Y + \epsilon_Y] = Cov[X + \epsilon_X] + Cov[Y + \epsilon_Y]$$
$$= Cov[X,Y] + Cov[X,\epsilon_Y] + Cov[Y,\epsilon_Y] + Cov[\epsilon_X + \epsilon_Y] = Cov[X,Y] + Cov[\epsilon_X, \epsilon_Y]$$
$$= Cov[X,Y]$$

Hence, $\rho_{\tilde{X},\tilde{Y}} = \frac{1}{1+b}\rho_{X,Y}$